\documentclass[11pt,twoside]{article}

\usepackage{asp2004}
\usepackage{epsf}
\usepackage{psfig}
\usepackage{lscape}
\usepackage{url}
\begin{document}
\title{Observational constraints on the evolutionary connection
  between PG~1159 stars and DO white dwarfs}   
\author{S.~D.~H\"ugelmeyer$^1$, S.~Dreizler$^1$, K.~Werner$^2$,
  J.~Krzesi\'nski$^{3,4}$,  A.~Nitta$^5$, and S.~J.~Kleinman$^6$}
\affil{$^1$Institut f\"ur Astrophysik, Universit\"at$\,$G\"ottingen,
  Friedrich-Hund-Platz~1, D-37077 G\"ottingen, Germany \\ $^2$Institut
  f\"ur Astronomie und Astrophysik, Universit\"at$\,$T\"ubingen, Sand~1,
  D-72076 T\"ubingen, Germany\\ $^3$New Mexico State University, Apache
  Point Observatory, 2001 Apache Point Road, P.O. Box 59, Sunspot, NM
  88349, USA\\ $^4$Mt. Suhora Observatory, Cracow Pedagogical
  University, ul. Podchorazych 2, 30-084 Cracow, Poland \\ $^5$Gemini
  Observatory, 670 N. A'Ohoku Place, Hilo, HI 96720, USA \\ $^6$Subaru
  Telescope, National Astronomical Observatory of Japan, 650 N. A'Ohoku
  Place, Hilo, HI 96720, USA}

\begin{abstract} 
The Sloan Digital Sky Survey has provided spectra of a large number of
new PG~1159 stars and DO white dwarfs. This increase in known hot
H-deficient compact objects significantly improves the statistics and
helps to investigate late stages of stellar evolution. We have
finished our analyses of nine PG~1159 stars and 23 DO white dwarfs by
means of detailed NLTE model atmospheres. From the optical SDSS
spectra, effective temperatures, surface gravities, and element
abundances are derived by using our new automated $\chi^2$--fitting in
order to place the observed objects in an evolutionary
context. Especially the connection between PG~1159 stars and DO white
dwarfs has been investigated.
\end{abstract}


\section{Introduction} 

PG~1159 stars are evolutionary transition objects between the hottest
post-AGB and WD phases. The Palomar Green (PG) survey put forth the
prototype of this spectroscopic class, PG~1159-035
\citep{1986ApJS...61..305G}. It shows a spectrum without detectable
hydrogen lines and is instead dominated by \ion{He}{ii} and highly
ionised carbon and oxygen lines. PG~1159 stars display a
characteristic broad absorption trough around 4670~\AA\ composed of
\ion{He}{ii} 4686~\AA\ and several \ion{C}{iv} lines, suggesting high
effective temperatures. Spectral analyses yield $T_{\rm eff} = 75\,000
- 200\,000$~K and surface gravities of $\log{g} = 5.5 - 8.0$
\citep{1991A&A...244..437W,1994A&A...286..463D,1996aeu..conf..229W}.
Of the 29 PG~1159 stars known prior to the SDSS, ten are low-gravity
\citep[subtype lgE, ][]{1992LNP...401..273W} stars, placing them in
the same Hertzsprung-Russell-Diagram (HRD) region as the hot central
stars of planetary nebulae (CSPNe), while the others are more compact
objects with surface gravities of WDs (subtype A or E). Due to their
rarity, the majority of the known PG~1159 stars were discovered in
large surveys \citep[Palomar Green, Hamburg Schmidt (HS),
][]{1995A&AS..111..195H}. The most recent and only discovery within
the last 10 years, besides those from the SDSS
\citep{2004A&A...424..657W}, was an object from the Hamburg ESO (HE)
survey \citep{1996A&AS..115..227W}. The SDSS thus offers a new and
currently unique opportunity to increase the number of known PG~1159
stars.

White dwarfs can be separated into two distinct spectroscopic classes,
DA and non-DA white dwarfs. The former show a pure hydrogen spectrum
and can be found over the entire WD cooling sequence. The latter fall
into three subclasses: DO ($45\,000~{\rm K} < T_{\rm eff} <
120\,000$~K), DB ($11\,000~{\rm K} < T_{\rm eff} < 30\,000$~K), and DC
/ DQ / DZ white dwarfs ($T_{\rm eff} < 11\,000$~K; pure continuum /
carbon / metal lines present). The classification of white dwarfs of
subtype DO and DB is determined by the ionisation balance of
\ion{He}{i} and \ion{He}{ii}. DO white dwarfs show a pure \ion{He}{ii}
spectrum at the hot end and a mixed \ion{He}{i/ii} spectrum at the
cool end. The transition to the cooler DB dwarfs, characterised by
pure \ion{He}{i} spectra, is interrupted by the so-called ``DB-gap''
\citep{1986ApJ...309..241L}. In the HRD region of white dwarfs with
$30\,000~{\rm K} < T_{\rm eff} < 45\,000$~K, no objects with
H-deficient atmospheres have been observed to date. However,
\citet{2006AJ....132..676E} describe over 25 objects taken from SDSS
DR4 data which likely fall in that temperature region. These new
objects might increase our knowledge of the spectral evolution of
He-rich white dwarfs considerably.

The region in the HRD occupied by the PG~1159 stars overlaps with that
of the DO white dwarfs. Therefore, it is assumed that gravitational
settling of the heavier elements in the atmosphere of the PG~1159
stars leads them to transition towards DO white dwarfs. Diffusion
calculations by \citet{2000A&A...359.1042U} support this
assumption. They show that the decrease of winds from PG~1159 stars
leads to a rapid settling of the CNO elements.

\section{Models and fitting}

For the PG~1159 stars in our sample, we calculated NLTE model
atmospheres with TMAP, the T\"ubingen NLTE Model Atmosphere Package
\citep{2003WSAM...W,2003WSAM...R}, using detailed
{H}\,--\,{He}\,--\,{C}\,--\,{N}\,--\,{O} model atoms. The model grid
ranges from $T_{\rm eff}= 55\,000 - 150\,000$~K and $\log{g} = 5.5 -
7.8$. The abundances are fixed to values ${\rm He}/{\rm H}=100$ and
${\rm C}/{\rm He}=0.01 - 0.11$ (in steps of 0.02), $0.20$, $0.30$, or
$0.60$ (number ratio). We have a nearly complete model grid with $T_{\rm
eff}=55\,000 - 110\,000$~K in steps of $5\,000$~K, $\log{g}=6.4 - 7.8$
in steps of $0.2$~dex, and ${\rm C}/{\rm He}=0.01-0.11$ in steps of
$0.02$. Complete coverage of the whole parameter space is not yet
available due to the high computational time required to compute the
model atmospheres. However, the majority of the analysed PG~1159 stars
are covered by our nearly complete model grid. Best-fit models for
PG~1159 star candidates were calculated with an oxygen abundance
following the typical PG~1159 abundance-scaling ratio ${\rm O}/{\rm C}
\approx {\rm C}/{\rm He}$. However, variations in the oxygen abundance
do not {produce} a significant effect on the other stellar
parameters. The nitrogen abundance is kept fixed at ${\rm N}/{\rm
He}=0.01$ by number which is a typical upper limit for PG~1159 stars.

Detailed {H}\,--\,{He} atomic models \citep{1996A&A...314..217D} were
used to calculate our NLTE model atmospheres for the DO white
dwarfs. The model grid ranges from $T_{\rm eff} = 40\,000 -
120\,000$~K in steps of $2\,500$~K, $\log{g}$ ranges from 7.0 to 8.4
in intervals of $0.2$~dex. The helium abundance is fixed to ${\rm
He}/{\rm H} = 99$.

In order to derive a best-fit model we applied a
$\chi^2$-analysis. The noise level $\sigma$ of the observed spectrum
has been determined using a Savitzky-Golay smoothing filter
\citep{1964AnaCh..36.1627S}. The observation was then normalised to
unity with the help of the normalised model spectrum. Given the
continuum points from the model, we fitted a third order polynomial
through the double-logarithmic flux-wavelength data of the
observation, which have a linear relation in good approximation. This
is due to the fact that the optical SDSS spectra of hot WDs are
dominated by the Rayleigh-Jeans tail of the flux distribution, which
behaves as $F_\lambda \sim \lambda^{-4}$. Dividing the spectrum by the
polynomial yields the normalised data which can now be compared to the
synthetic spectrum. We have done this consistently for all spectra in
order to reduce errors introduced by the individual normalisation of
the observation. For the determination of the $\chi^2$ value we
selected lines and regions that are essential for the quality of the
fit.

\section{Results}

In our two spectral analyses papers
\citep{2005A&A...442..309H,2006A&A...454..617H}, we have presented the
atmospheric parameters of our programme stars and plots of observed
spectra and best-fit models. Here we now want to compare SDSS results
to previous work and go more into detail on the transition of PG~1159
stars towards DO white dwarfs.

\subsection*{PG~1159 stars}

Our analyses added eight new PG~1159 stars to the sample of 29 known
objects of this spectral type. Comparing the atmospheric parameters of
the new SDSS objects to the literature values of previously known
PG~1159 stars \citep[see \emph{e.g.}\ the review
of][]{2006PASP..118..183W}, it becomes obvious that our results
diverge from earlier ones. While the prototype PG~1159$-$035 yields
$T_{\rm eff}=140\,000$~K, $\log{g}=7.0$, and ${\rm C}/{\rm He}=0.60$,
and the other PG~1159 stars scatter around these values, seven out of
the eight new SDSS PG~1159 stars have $T_{\rm eff} \approx
100\,000$~K, $\log{g} \approx 7.5$, and a carbon abundance of ${\rm
C}/{\rm He}=0.05-0.30$. The only new object that deviates from this
average SDSS PG~1159 star is SDSS 001651.42$-$011329.3
\citep{2005A&A...442..309H} with $T_{\rm eff}=120\,000$~K,
$\log{g}=5.5$, and ${\rm C}/{\rm He}=0.30$ and is therefore of subtype
lgE. The reason why most of the SDSS PG~1159 stars are found in the
low temperature and high surface gravity part of the PG~1159 star domain is
easily explained by the fact that the evolutionary time scales are
much longer in this part of the post-AGB sequence than in the earlier
low gravity part. Therefore, the SDSS objects show what we would
expect from the theory. We should find even more objects in this late
stage of PG~1159 star evolution, however, the SDSS is only complete to
$\sim 30\%$ \citep{2006astro.ph..6700E} and therefore misses more than half of
these objects. Since 14 PG1159s have been found because they are CSPNs and
the three hottest objects have been detected in X-ray surveys, the
fact that preferably hotter and less compact -- \emph{i.e.}\ less
evolved -- PG~1159 stars have been discovered before the large SDSS
survey is no surprise.

One object from the SDSS sample shows lines of ultra-high excitation
ions (UHEI) such as \ion{C}{vi}, \ion{N}{vi}, \ion{N}{vii},
\ion{O}{vii}, \ion{O}{viii}, and even \ion{Ne}{x}. It is the first
PG~1159 star to do so among several known DO and DAO white dwarfs
presented in \citet{1995A&A...303L..53D} and
\citet{1995A&A...293L..75W}. Since the UHEI lines are blue shifted and
since they have an extreme triangular shape it is believed that they
originate in a hot and fast wind while the underlying star shows
He\,II and C\,IV lines that are too strong to be fitted with our model
spectra. An explanation for this phenomenon has not been found to
date.

\subsection*{DO white dwarfs}

In contrast to the SDSS PG~1159 stars, the analysed DO white dwarfs
from this survey cover the whole known temperature range and even
extend this to a new lowest value of $T_{\rm eff}=40\,000$~K placing
it in the ``DB-gap''. This is in agreement with the work of
\citet{2006AJ....132..676E} who find four DO white dwarfs from the
SDSS that are close to the suggested lower ``DB-gap'' temperature
boundary of $T_{\rm eff}=45\,000$~K and six DB just below $T_{\rm
eff}=40\,000$~K. Three of these four DOs have also been analysed in
\citet{2005A&A...442..309H,2006A&A...454..617H} and we found
temperatures agreeing with those of \citet{2006AJ....132..676E} by
about $\sim 2\,500$~K using our NLTE models while the latter authors
employed LTE synthetic spectra.

Two DOs from our sample show UHEI lines \citep[first classified
  by][]{2004A&A...417.1093K} and three have M-star features in their
spectra. Of the latter three, we believe that two are part of a
physical binary system.

Even though the SDSS DOs cover the expected parameter ranges for this
type of object, the mass distribution is shifted towards higher masses
by $0.10~M_{\sun}$ to $\overline{M}=0.69~M_{\sun}$ (not taking binary
WDs into account) compared to earlier analyses
\citep{1996A&A...314..217D}. This effect can also be seen by the
relatively high $\log{g}$ values of the SDSS DOs in
Fig.~\ref{fig:pgdo}. We believe that the increase in $\log{g}$ and
mass, respectively, most likely results from an incorrect calibration
of SDSS hot WD spectra since \citet{2004ApJ...607..426K} also observe a
deviation from expected values by comparing fitted DA parameters from
SDSS spectra to literature values of the observed stars. Their work
shows an overestimation of effective temperature for stars with
$T_{\rm eff} \ga 30\,000$~K which is compensated by their fitting
routines by also increasing $\log{g}$. This problem is discussed in
more detail by \citet{2006astro.ph..6700E}.

\section{Discussion}

\begin{figure}
  \plotone{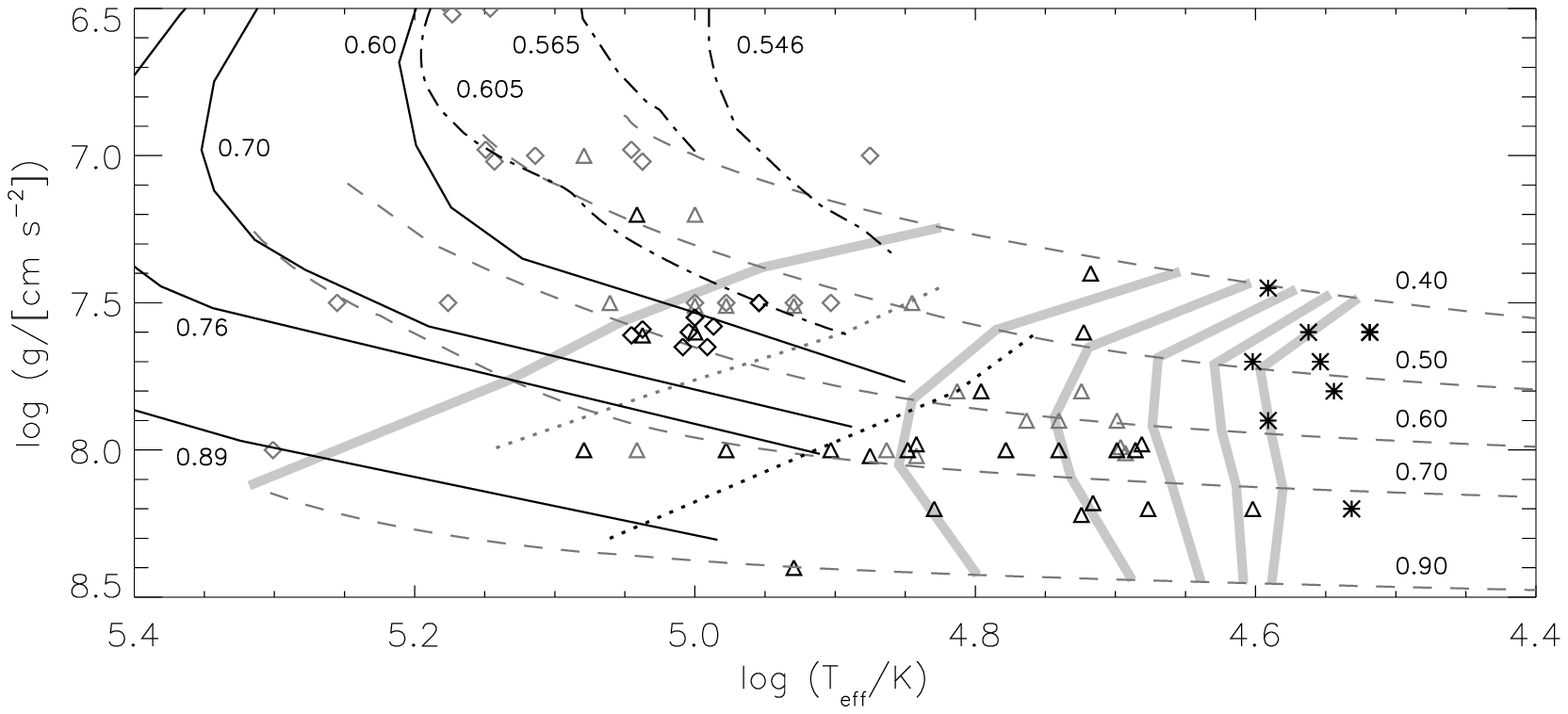}
  \caption{PG~1159 stars (diamonds) and DO white dwarfs (triangles)
    compared to evolutionary tracks from \citet{1995A&A...299..755B}
    and \citet{1983ApJ...272..708S} (dashed-dotted lines),
    \citet[][solid lines]{1986ApJ...307..659W}, and \citet[][dashed
    lines]{1995LNP...443...41W}. The black symbols come from our
    analyses of SDSS spectra and the grey ones from literature values
    \citep{2006PASP..118..183W,1996A&A...314..217D}. The black
    asterisks are hot DB white dwarfs in the ``DB-gap''
    \citep{2006AJ....132..676E}. The dotted lines are the wind limits for
    PG~1159 stars calculated by \citet{2000A&A...359.1042U}, where the
    grey line represents the limit calculated for a mass loss rate ten
    times lower than for the black line. The thick
    grey lines are equidistant time marks of $10^6$~years. Labels are
    given in $M_{\sun}$. \label{fig:pgdo}}
\end{figure}

We have placed the PG~1159 stars and DO white dwarfs from the SDSS and
previous analyses in an evolutionary context depicted in
Fig.~\ref{fig:pgdo}. Along with post-AGB and WD tracks, we have
plotted equidistant time marks of $10^6$ years. Despite the increasing
number of hot H-deficient (pre-)WDs, we are still dealing with low
number statistics. The time marks can though give a rough estimate on
the age distribution of He-rich WDs. We can see that the age bins above
$T_{\rm eff} > 45\,000$~K are quite equally populated if we only
regard DO white dwarfs. The new results of \citet{2006AJ....132..676E}
show that there is a continuation of H-deficient white dwarfs towards
lower temperatures than $T_{\rm eff} = 45\,000$~K. These hot DBs on
average have lower surface gravities than the hotter DOs. If this is a
real effect or a result of the application of LTE models to these hot
objects remains to be seen.

Concerning the transition between PG~1159 stars and DO white dwarfs,
we have plotted the wind limits for PG~1159 stars calculated by
\citet{2000A&A...359.1042U} in Fig.~\ref{fig:pgdo}. Because of
different initial compositions, PG~1159 stars have different mass loss
rates assuming that $\dot{M}$ depends on the composition. Therefore
the rapid depletion of the CNO elements predicted by
\citet{2000A&A...359.1042U} takes place at different positions in the
HRD. There are still a few DOs beyond either of the wind limits but
none of the PG~1159 stars from the SDSS, even though they are quite
evolved compared to previous analyses, has crossed the wind
limit. Therefore, we can say that the wind limit is a strict lower
limit for PG~1159 star -- DO white dwarf transitions.

The low carbon abundances observed in five of the eight new PG~1159
stars from the SDSS \citep[${\rm C/He}=0.03-0.05$ by
number,][]{2006A&A...454..617H} can also be explained by the diffusion
calculations of \citet{2000A&A...359.1042U}. Even though the
gravitational settling of the CNO elements sets in quite abrupt, the
metal abundances are already reduced by a factor of two compared to
the initial chemical composition at temperatures of the low-abundance
SDSS PG~1159 stars. Therefore, these stars can already be considered
transition objects.

\acknowledgements S.D.H. would like to thank the Royal Astronomical
Society and the Berliner-Ungewitter-Stiftung for their generous
financial support.


\end{document}